\documentclass[aps,prx,reprint,superscriptaddress]{revtex4-2}
\usepackage[T1]{fontenc}
\usepackage[utf8]{inputenc}
\usepackage{times,amsmath,amssymb,graphicx,enumitem}
\usepackage[dvipsnames]{xcolor}
\usepackage[colorlinks=true,allcolors=Blue]{hyperref}

\newcommand{\PHS}{ZnCu$_{3}$(OH)$_{6}$Cl$_{2}$}
\newcommand{\DHS}{ZnCu$_{3}$(OD)$_{6}$Cl$_{2}$}
\newcommand{\Broch}{ZnCu$_{3}$(OH)$_{6}$SO$_{4}$}
\newcommand{\VOF}{[NH$_{4}$]$_{2}$[C$_{7}$H$_{14}$N][V$_{7}$O$_{6}$F$_{18}$]}
\newcommand{\ET}{$\kappa$-(BEDT-TTF)$_{2}$Cu$_{2}$(CN)$_{3}$}
\newcommand{\dmit}{EtMe$_{3}$Sb[Pd(dmit)$_{2}$]$_{2}$}

\begin{document}

\title{Specific Heat of the Kagome Antiferromagnet Herbertsmithite in High Magnetic Fields}

\author{Quentin Barthélemy}
\affiliation{Université Paris-Saclay, CNRS, Laboratoire de Physique des Solides, 91405, Orsay, France}

\author{Albin Demuer}
\affiliation{LNCMI-EMFL, CNRS UPR3228, Univ. Grenoble Alpes, Univ. Toulouse, Univ. Toulouse 3, INSA-T, Grenoble and Toulouse, France}

\author{Christophe Marcenat}
\affiliation{Univ. Grenoble Alpes, CEA, Grenoble INP, IRIG, PHELIQS, 38000, Grenoble, France}

\author{Thierry Klein}
\affiliation{Université Grenoble Alpes, CNRS, Grenoble INP, Institut Néel, 38000, Grenoble, France}

\author{Bernard Bernu}
\affiliation{Sorbonne Université, CNRS, Laboratoire de Physique Théorique de la Matière Condensée, 75005, Paris, France}

\author{Laura Messio}
\affiliation{Sorbonne Université, CNRS, Laboratoire de Physique Théorique de la Matière Condensée, 75005, Paris, France}
\affiliation{Institut Universitaire de France, 75005, Paris, France}

\author{Matias Velázquez}
\affiliation{Université Grenoble Alpes, CNRS, Grenoble INP, SIMAP, 38000, Grenoble, France}

\author{Edwin Kermarrec}
\affiliation{Université Paris-Saclay, CNRS, Laboratoire de Physique des Solides, 91405, Orsay, France}

\author{Fabrice Bert}
\affiliation{Université Paris-Saclay, CNRS, Laboratoire de Physique des Solides, 91405, Orsay, France}

\author{Philippe Mendels}
\email{philippe.mendels@universite-paris-saclay.fr}
\affiliation{Université Paris-Saclay, CNRS, Laboratoire de Physique des Solides, 91405, Orsay, France}

\date{\today}

\begin{abstract}
Measuring the specific heat of herbertsmithite single crystals in high magnetic fields (up to $34$ T) allows us to isolate the low-temperature kagome contribution while shifting away extrinsic Schottky-like contributions. The kagome contribution follows an original power law $C_{p}(T\rightarrow0)\propto T^{\alpha}$ with $\alpha\sim1.5$ and is found field-independent between $28$ and $34$~T for temperatures $1\leq T\leq4$~K. These are serious constrains when it comes to replication using low-temperature extrapolations of high-temperature series expansions. We manage to reproduce the experimental observations if about $10$~\% of the kagome sites do not contribute. Between $0$ and $34$~T, the computed specific heat has a minute field dependence then supporting an algebraic temperature dependence in zero field, typical of a critical spin liquid ground state. The need for an effective dilution of the kagome planes is discussed and is likely linked to the presence of copper ions on the interplane zinc sites. At very low temperatures and moderate fields, we also report some small field-induced anomalies in the total specific heat and start to elaborate a phase diagram.
\end{abstract}

\maketitle

\section{\label{sec:intro}Introduction}

Herbertsmithite \PHS\ plays a leading role in the investigation of quantum spin liquid states, as the most studied representative material of the quantum kagome Heisenberg antiferromagnet model (QKHA model)~\cite{Mendels2016,Norman2016}. In herbertsmithite, the magnetic Cu$^{2+}$ with spin $S=1/2$ decorate the vertices of structurally perfect kagome planes. The strong in-plane antiferromagnetic coupling $J\sim190$~K between nearest neighbors is known to exceed by far all other couplings between further in-plane or interplane neighbors~\cite{Jeschke2013}. This leads to a highly bidimensional and optimally frustrated quantum magnet which indeed does not order nor freeze down to the lowest temperatures reached in experiments ($20$~mK $\sim J/9000$)~\cite{Mendels2007,Mendels2016} and features a continuum of fractionalized excitations~\cite{Han2012-bis,Han2016}.

The nature of the ground state of the QKHA model is still highly debated on the theory side~\cite{Hermele2008,Iqbal2014,Liao2017,He2017,Zhu2018,Lee2018,Chen2018,Hering2019,Jiang2019,Lauchli2019} while experimental investigations have not yet been able to discriminate unambiguously between the proposed scenarios. Indeed, inevitable deviations to the pure QKHA model call for caution when comparing measurements performed on real quantum kagome antiferromagnets to theoretical predictions. In herbertsmithite, the lack of inversion symmetry within copper-copper bonds is responsible for an appreciable out-of-plane Dzyaloshinsky-Moriya anisotropy $D_{z}$ in addition to a possible anisotropy of the symmetric exchange tensor. The combination of both effects is believed to matter as $0.06(2)J$ at most~\cite{Zorko2008,ElShawish2010,Han2012}. Besides, a fraction $f\sim20$~\% of interplane zinc sites is considered to be occupied by copper ions. The low-temperature Curie-like susceptibility $\chi^{\mathrm{Imp}}(T\rightarrow0)\propto f/T$ makes them appear as simple paramagnetic objects that can be treated independently from the kagome physics. Nevertheless, the spatial extent of these magnetic defects has remained elusive and there is some debate whether perfectly stoichiometric \PHS\ actually exists~\cite{Smaha2020}.

These magnetic defects are highly troublesome for bulk-averaged thermodynamic measurements because their response dominates the kagome behavior at low temperatures and low magnetic fields. Thus, local probes have proven to be essential to gain information on the dispersion relation and density of states of elementary excitations, which are observables of interest to determine the precise nature of the ground state~\cite{Olariu2008,Fu2015,Khuntia2020}. In a recent $^{17}$O NMR study on a single crystal, Khuntia \emph{et al.} were able to contrast and filter out the defect contributions in shift and spin-lattice relaxation rate measurements~\cite{Khuntia2020}. At low temperatures, both quantities exhibit a nearly linear trend. These observations are robust evidences against the presence of a gap in the excitation spectrum. Despite sizeable error bars, it enabled a first quantitative comparison with series expansions for the static susceptibility~\cite{Bernu2015}. Since the density of states appears to be strongly reduced (if not vanishing) in the $T\rightarrow0$ limit, the low-energy spectrum might feature Dirac nodes rather than a well defined Fermi surface, in the case of fractional $S=1/2$ spinon excitations. This conclusion is in overall agreement with the latest results obtained for the QKHA model with several numerical techniques~\cite{Hermele2008,Iqbal2014,Liao2017,He2017,Zhu2018,Lee2018,Chen2018,Hering2019,Jiang2019,Lauchli2019}. Other unbiased low-temperature measurements that could be easily compared to theory are now crucial to move forward on the exact nature of the emergent gapless quasiparticles.

In this context, the low-temperature specific heat $C_{p}$ is of great interest because it is a direct probe of the low-energy density of states. Clear-cut differences between models can be expected in the $T\rightarrow0$ limit. For instance, free Dirac fermions lead to the graphene-like $C_{p}(T,B=0)\propto T^{2}$ while a spinon Fermi sea yields the metal-like $C_{p}(T)\propto T$, although fluctuations of emergent gauge fields may renormalize these mean-field behaviors~\cite{Kim1997}. Unfortunately, all specific heat studies performed to-date have suffered from the difficulty to get rid of external contributions aside from that of the kagome planes. Indeed, on the low-temperature range of interest, the specific heat displays a field dependence similar to the Schottky anomaly of two-level systems :
\begin{equation}\label{eqimp}
C_{p}^{\mathrm{Imp}}(T,\Delta)=f\frac{\mathcal{N}_{\mathrm{A}}k_{\mathrm{B}}\Delta^{2}\exp{(\Delta/T)}}{T^{2}\left[1+\exp{(\Delta/T)}\right]^{2}},
\end{equation}
where the fraction of these two-level systems per mole of formula is found to be corresponding to the fraction $f$ of zinc sites occupied by copper ions~\cite{deVries2008} and $\Delta$ is their energy gap (in Kelvin units). This gap $\Delta$ has been found to follow the standard Zeeman splitting $g\mu_{\mathrm{B}}B/k_{\mathrm{B}}$ with $g=2.2$ for fields $B>2$~T~\cite{deVries2008}. As the temperature associated with the maximum of $C_{p}^{\mathrm{Imp}}$ increases linearly with the applied field, intense fields may be used to make the low-temperature contribution $C_{p}^{\mathrm{Imp}}(T)\sim\exp{(-\Delta/T)}/T^{2}$ negligible. In previous studies, the fields have never been high enough (limited to $18$~T) to extract unambiguously the kagome contribution on a reasonably wide temperature range for robust interpretations~\cite{Helton2007,deVries2008,Han2012,Han2014}.

Here, we report specific heat measurements on herbertsmithite single crystals for temperatures $0.7\leq T\leq8$~K and fields up to $34$~T.

\section{\label{sec:details}Experimental details}

For this study, we used a single crystal of protonated herbertsmithite \PHS\ (PHS) and a single crystal of deuterated herbertsmithite \DHS\ (DHS) that were prepared using the aqueous growth procedure described in Ref.~\cite{Velazquez2020}. From SQUID macroscopic susceptibility measurements~\cite{sup}, the ratio of the two low-temperature Curie constants is $q\equiv f_{\mathrm{DHS}}/f_{\mathrm{PHS}}=1.3(1)$. This is related to the different temperatures used in the growth process~\cite{Velazquez2020}. The DHS single crystal is thus believed to harbour a slightly larger amount of nearly free magnetic entities. The specific heat measurements were carried out by means of the original ac modulation technique introduced in Ref.~\cite{Michon2019} on a slab-like piece from each crystal, with a mass of $87(2)$~$\mu$g for the PHS sample and a mass of $286(2)$~$\mu$g for the DHS sample~\cite{sup}. About $20(5)$~$\mu$g of Apiezon grease was used to attach the crystals to the Cernox resistive chips. The empty chips (with grease) were first measured in order to subtract the addenda specific heat~\cite{sup}. No special care was taken regarding the sample orientation with respect to the field direction. Indeed, a previous study on single crystals reported a marginal anisotropy, which is below our absolute accuracy $\Delta C_{p}/C_{p}\sim5$~\% for $T>1$~K~\cite{Han2012}.

\section{\label{sec:raw}Direct measurement of the kagome contribution $C_{p}^{\mathrm{Kago}}$}

\begin{figure}\centering
\includegraphics[width=\columnwidth]{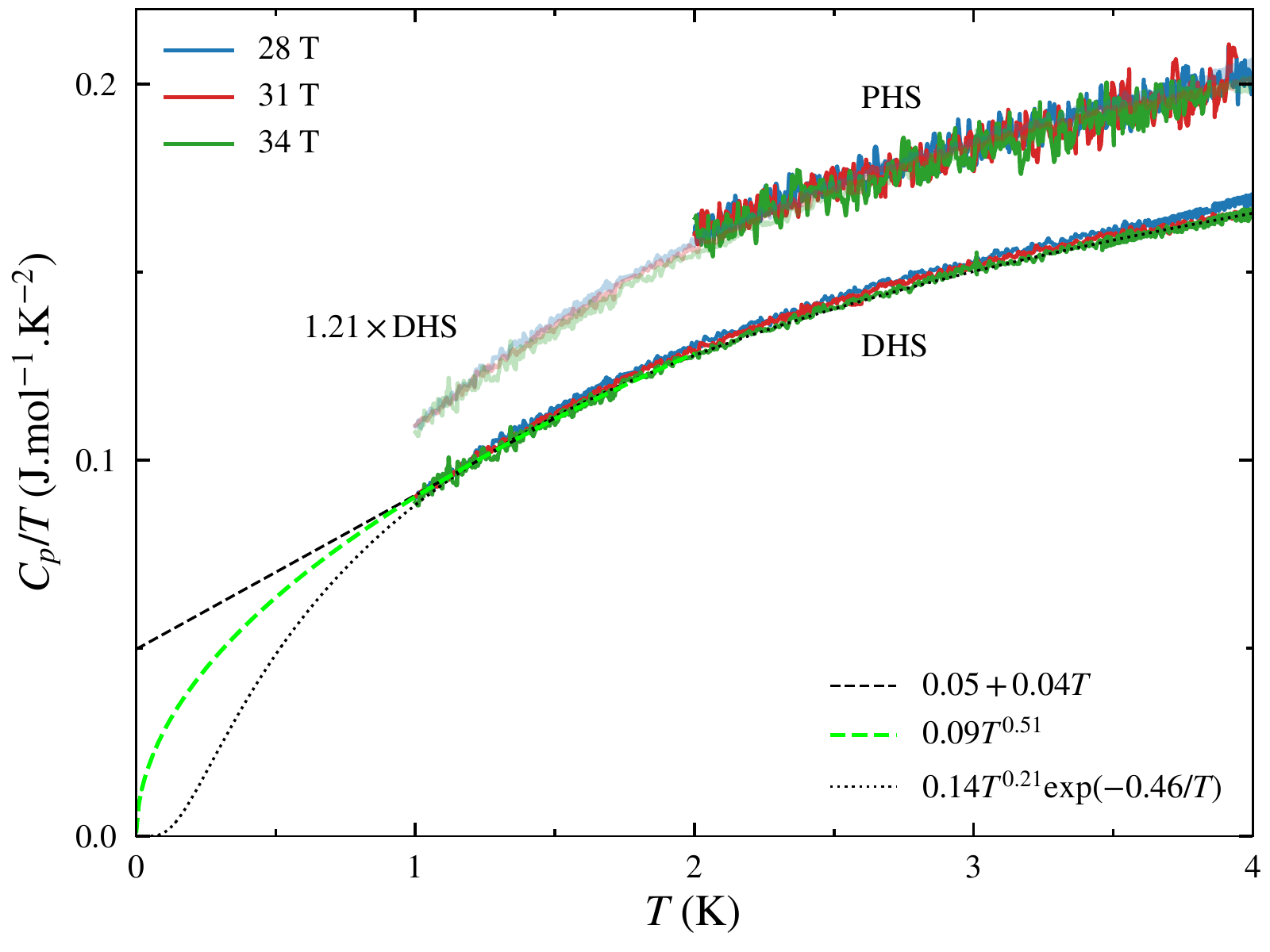}
\caption{\label{fig:main1}
Temperature scans obtained at $28$, $31$ and $34$~T for temperatures $T\leq4$~K for the PHS and DHS single crystals. The phononic and Schottky-type contributions are negligible in the considered temperature and field ranges. Hence, the kagome contribution is directly measured and found to be field-independent. The PHS and DHS data are found to scale by a factor 1.21, see the lines in transparency. The dashed lines indicate possible extrapolations to zero temperature, as detailed in the text.}
\end{figure}

\begin{figure}\centering
\includegraphics[width=\columnwidth]{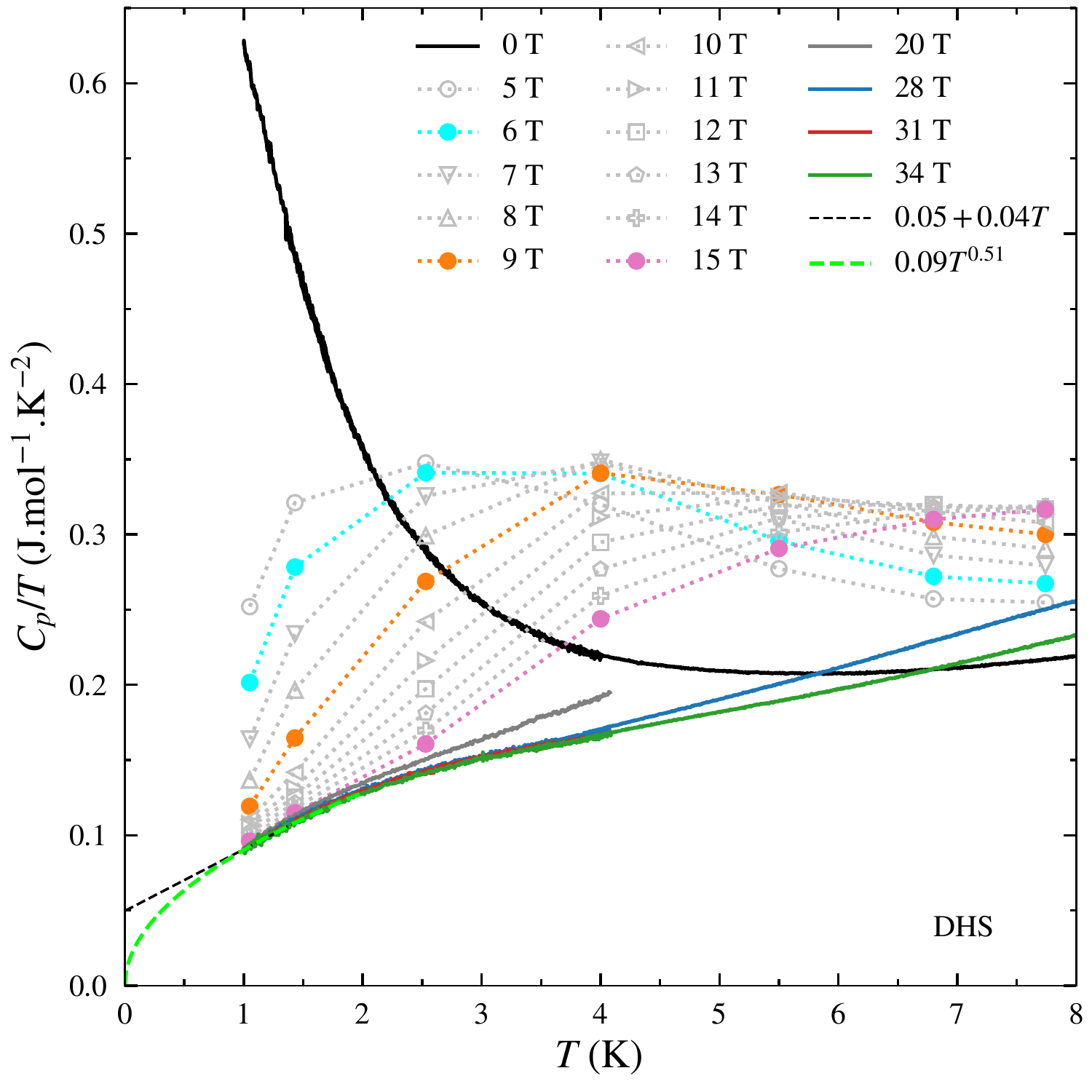}
\caption{\label{fig:main2} $C_{p}/T$ data obtained at $16$ selected field values for temperatures $1\leq T\leq8$~K on the DHS single crystal. The discontinuous datasets are extracted from field scans at fixed temperatures. Some of the extrapolations from Fig.~\ref{fig:main1} are reproduced here.}
\end{figure}

We begin with a description of the specific heat measurements at the highest fields available in our experiments. The total specific heat can be decomposed into three contributions:
\begin{equation}
C_{p}=C_{p}^{\mathrm{Kago}}+C_{p}^{\mathrm{Schot}}+C_{p}^{\mathrm{Phon}}
\end{equation}
where the Schottky-like contribution $C_{p}^{\mathrm{Schot}}$, arising from the nearly free magnetic entities~\footnote{We do not detect the nuclear Schottky contributions on the range $1-4$~K, likely because of too long relaxation times.}, and the lattice contribution $C_{p}^{\mathrm{Phon}}$ add up to the kagome contribution of interest $C_{p}^{\mathrm{Kago}}$. For both samples, we show in Fig.~\ref{fig:main1} the temperature scans obtained at $28$, $31$ and $34$~T for $1\leq T\leq4$~K. The estimated Schottky-like and lattice contributions are negligibly small on this temperature range ($C_{p}^{\mathrm{Phon}}/T<12.5$~mJ.mol$^{-1}$.K$^{-2}$ at 4~K and, for DHS, $C_{p}^{\mathrm{Schot}}/T<1.8$~mJ.mol$^{-1}$.K$^{-2}$~\cite{sup}), implying that we are directly probing the kagome contribution $C_{p}^{\mathrm{Kago}}$. We show in Fig.~\ref{fig:main2} the complete evolution of $C_{p}/T$ from $0$ to $34$~T on a broader temperature range for the DHS sample. As expected, the peak associated with the Schottky-like contribution clearly shifts towards high temperatures as the field increases and only a minute contamination remains below $4$~K for fields $B\geq28$~T. On the $T\leq4$~K range and in these high fields, $C_{p}^{\mathrm{Kago}}$ is found to be field-independent in both samples.

Despite a similar $C_{p}/T$ evolution, there is a significant difference in magnitude between the PHS and DHS samples. We find that a simple scaling of the DHS data by a factor $1.21$ reproduces the observations on the PHS sample, as shown in Fig.~\ref{fig:main1}. This 1.21 factor is remarkably close to the ratio of the low temperature Curie constants $q=1.3(1)$. Hence the difference likely originates from the distinct impurity concentrations.

We now turn to various extrapolations of our data in the $T\rightarrow0$ limit based on existing models. A fit to $C_{p}(T)/T=aT^{\eta}\exp{(-\Delta/T)}$, see Fig.~\ref{fig:main1}, yields $\eta=0.21(5)$ and a gap $\Delta=0.5(1)$~K~$\sim J/360$, that is tiny compared to the estimate for a gapped spin liquid: $J/20\lesssim\Delta\lesssim J/10$~\cite{Depenbrock2012}. Sticking now to a gapless spin liquid, a fit in the range $[1,1.5]$~K to the empirical law $C_{p}(T)/T=\gamma+aT$ is used to provide an upper bound for any spinon ``Sommerfeld coefficient'' $\gamma$, see Fig.~\ref{fig:main1}. We obtain $\gamma=50(8)$~mJ.mol$^{-1}$.K$^{-2}$. In comparison to spin liquid candidates with a suggested spinon Fermi surface, this is much lower than for the kagome antiferromagnets \Broch~\cite{Li2014} and \VOF~\cite{LClark2013} ($200\lesssim\gamma\lesssim400$~mJ.mol$^{-1}$.K$^{-2}$) but higher than for the triangular organic salts \ET~\cite{Yamashita2008} and \dmit~\cite{Yamashita2011} ($\gamma\sim20$~mJ.mol$^{-1}$.K$^{-2}$). In the case of a critical spin liquid, the best fit in the range $[1,2]$~K is obtained using $C_{p}(T)/T=aT^{\rho}$ with a non-trivial power $\rho=0.51(4)$, see Fig.~\ref{fig:main1}.

\section{\label{sec:htse}Analysis with the HTSE$+s(e)$ method}

\begin{figure}\centering
\includegraphics[width=\columnwidth]{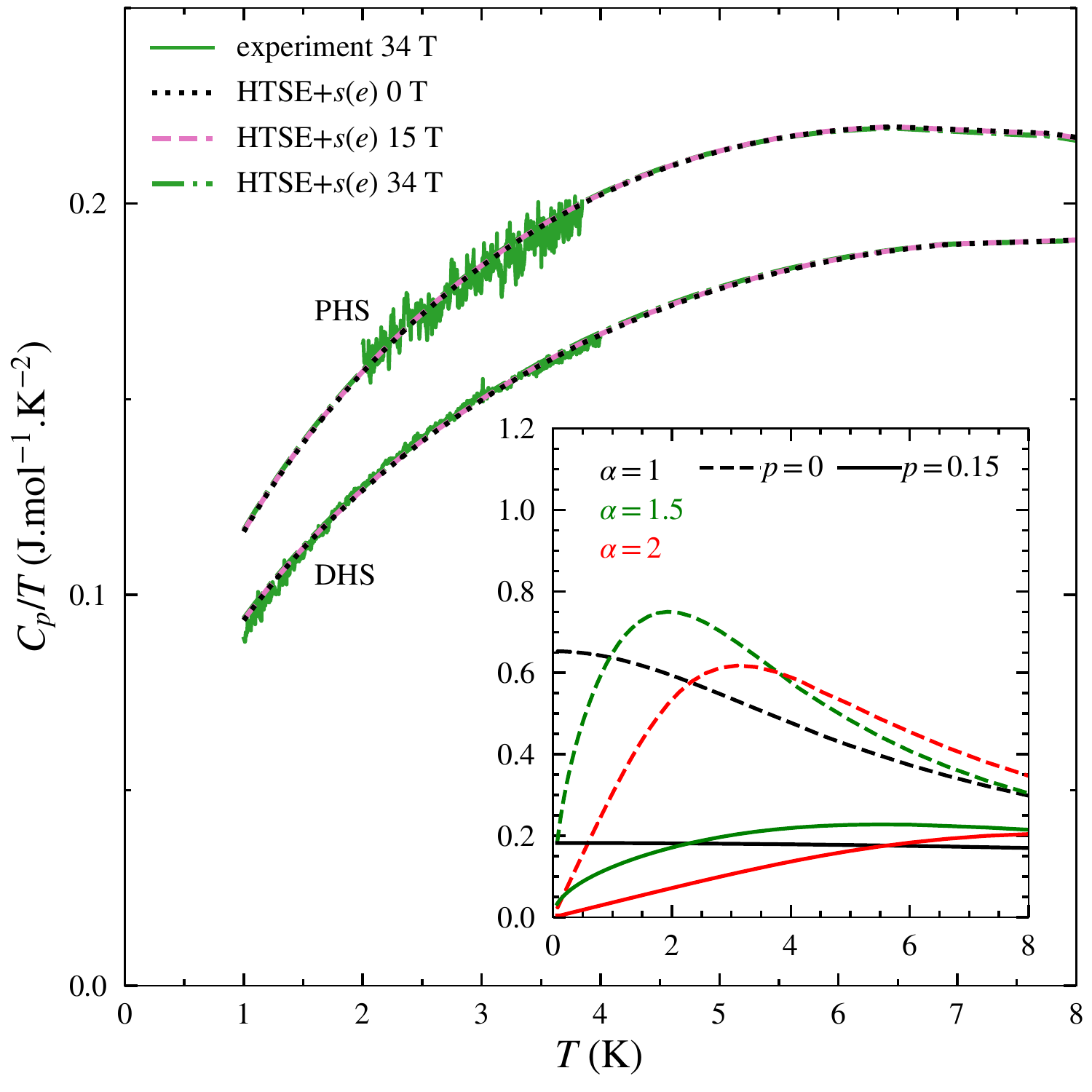}
\caption{\label{fig:main3}
$C_{p}/T$ data at $34$~T compared to the HTSE$+s(e)$ results which are field independent from $0$ to $34$~T. In the inset, we show the drastic impact of $\alpha$ and $p$ values on the shape and magnitude of the specific heat curves. The experimental data are only reproduced if we use a power $\alpha\sim1.5$ and impose a dilution $p\sim10-15$~\%.}
\end{figure}

The extrapolation scheme for high-temperature series expansions based on the entropy method (HTSE$+s(e)$ method) allows to compute the thermodynamic quantities corresponding to a spin hamiltonian down to zero temperature~\cite{Bernu2001,Misguich2005,Bernu2015,Bernu2020,sup}. The extrapolation is constrained by imposing the entropy behavior in the $T\rightarrow0$ limit and the zero temperature energy $e_{0}$. The latter is determined, for each set of fixed parameters of the hamiltonian, by looking for the maximum number of coinciding Padé approximants of the high temperature series expansion (HTSE) of an auxiliary function~\cite{Bernu2020}.

Here we use the QKHA model with $J=190$~K in an external magnetic field $B$ and with a dilution $p$, for which the HTSE in $1/T$ are known up to order 20~\cite{Bernu2020}. As shown in Fig.~\ref{fig:main3} (main and inset), the low value of the specific heat and its low T evolution can be reproduced if (i) a gapless $C_{p}\propto T^{\alpha}$ with $\alpha \sim 1.5$ is considered in the $T\rightarrow0$ limit, as already suggested by the power law fit, see Fig.~\ref{fig:main1}, (ii) a dilution $p \sim 0.1$ of the magnetic network is introduced. Other perturbations, like exchange anisotropies, may be considered~\cite{Bernu2020} but they have a weaker effect as compared to dilution. In the computation, the missing spins in the kagome planes are accounted statistically: each cluster of $n$ sites entering in the evaluation of the series expansions takes a weight $(1-p)^{n}$.

Regardless of any match to our experimental data, when a gapless $C_{p}(T)\propto T^{\alpha}$ ($1\leq\alpha\leq2$) is considered in the $T\rightarrow0$ limit, we find that the evolution of the ground state energy with the applied field is compatible with the expression:
\begin{equation}\label{eq:energy}
e_{0}(B,p)=e_{0}(B=0,p)-A(p) B^2
\end{equation}
Therefore, we used this ansatz with $p$, $e_{0}(B=0,p)$ and $A(p)$ as parameters~\cite{sup} to fit the experimental specific heat results. For $\alpha=1.45$, we find good fits to both PHS and DHS data and reproduce the observed field independence of the specific heat between 28~T and 34~T in the temperature range 1-4~K. The corresponding optimal dilution parameters are $p=0.11$ and $p=0.135$ for the PHS and DHS sample respectively. Interestingly, the ratio of these dilution parameters matches the ratio of the low-temperature Curie constants $q=1.3(1)$. This calls for a discussion of the origin of the effective dilution of the kagome planes needed to reproduce the specific heat data, see Sec.~\ref{sec:discussion}.

Finally, we assume that the parameters $\alpha$ and $p$ are field-independent and we compute the specific heat down to zero field for both PHS and DHS, see Fig.~\ref{fig:main3}. The computed specific heat has hardly any field dependence for temperatures $T\leq8$~K and for fields $B\leq34$~T.

\section{\label{sec:gaussian}Extraction and subtraction of the Schottky-like contribution $C_{p}^{\mathrm{Schot}}$}

\begin{figure}\centering
\includegraphics[width=\columnwidth]{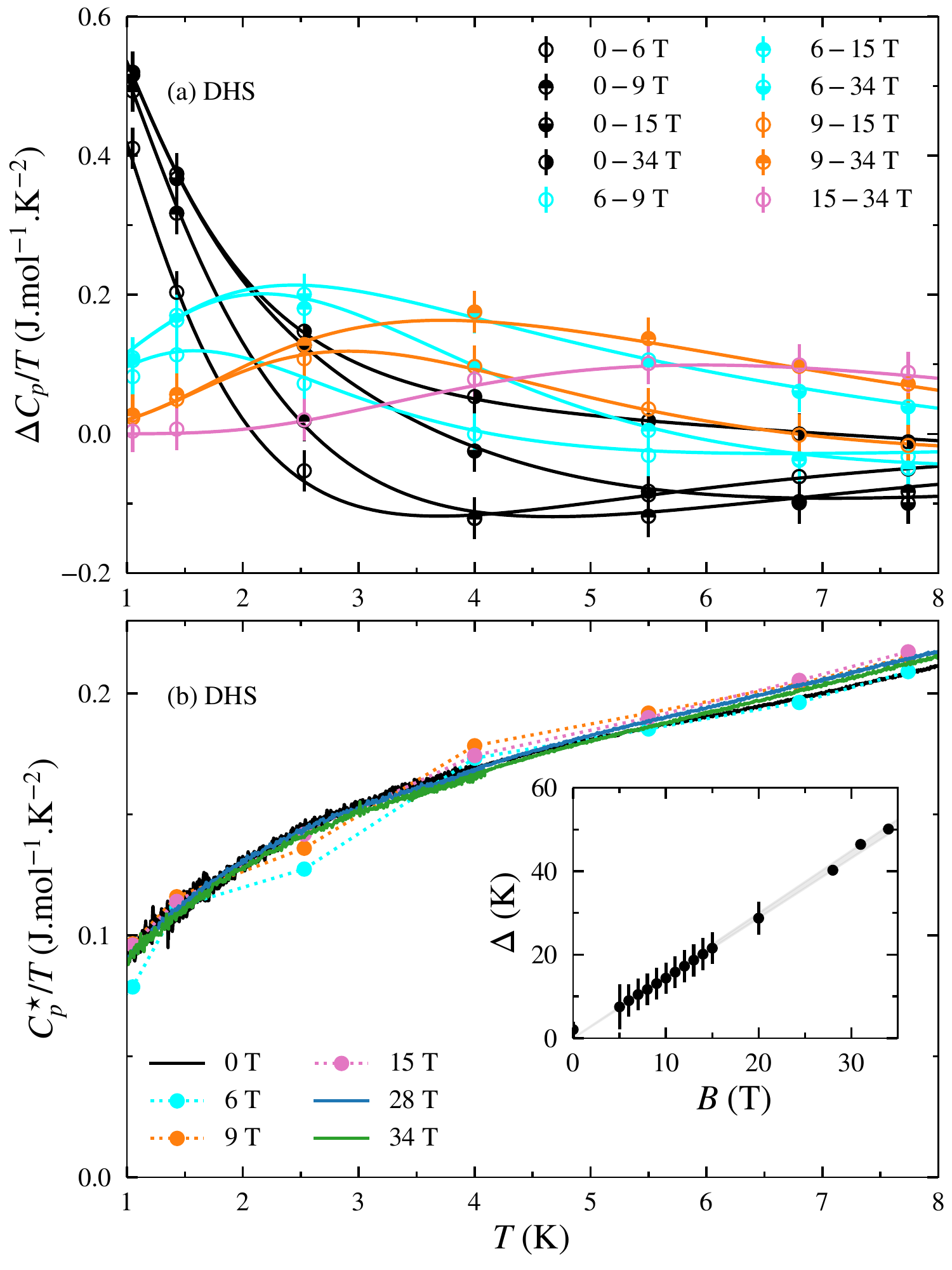}
\caption{\label{fig:main4}(a) We show $10$ among the $120$ differences $\Delta C_{p}/T$ allowed by the $16$ datasets of Fig.~\ref{fig:main2} (DHS single crystal only) and the corresponding fitted curves (solid lines) using the phenomenological model of Eq.~\ref{eq:dif}. (b) Once the estimated Schottky-type contributions are subtracted, we get the field-independent $C_{p}^{\star}/T=(C_{p}^{\mathrm{Kago}}+C_{p}^{\mathrm{Phon}})/T$. Inset: evolution of the gap $\Delta$, obtained from the latter fits, as a function of the field. The errorbars stand for the needed gaussian broadening $\sigma$.}
\end{figure}

In this section, we extend our analysis of the experimental data to temperature and field ranges for which the Schottky-like contribution becomes non-negligible and has to be subtracted. Motivated by the absence of field-dependence of the kagome specific heat discussed above, we propose a similar approach as developed in Ref.~\cite{deVries2008}.

We assume that the kagome contribution $C_{p}^{\mathrm{Kago}}$ is field-independent on the whole field range up to $34$~T. The Schottky-like field-dependent part of the total specific heat $C_{p}$ is singled out through differences between temperature scans obtained at fixed fields. We use $5$ parameters to describe the difference $\Delta C_{p}^{1,2}/T$ between the two datasets obtained at fields $B_{1}$ and $B_{2}$:
\begin{equation}\label{eq:dif}
\frac{\Delta C_{p}^{1,2}}{T}\equiv\frac{f}{T}\left[\overline{C_{p}^{\mathrm{Imp}}}(T,\Delta_{1},\sigma_{1})-\overline{C_{p}^{\mathrm{Imp}}}(T,\Delta_{2},\sigma_{2})\right],
\end{equation}
where
\begin{equation}\label{eq:gaussian}
\overline{C_{p}^{\mathrm{Imp}}}(T,\Delta,\sigma)\equiv\int C_{p}^{\mathrm{Imp}}(T,\tilde{\Delta})\rho_{\Delta,\sigma}(\tilde{\Delta})d\tilde{\Delta}.
\end{equation}
In the latter, we had to introduce  a gaussian distribution $\rho_{\Delta,\sigma}$ with mean $\Delta$ and standard deviation $\sigma$, and the integral is evaluated over $[\Delta-6\sigma,\Delta+6\sigma]\cap\mathbb{R}^{+}$. To achieve convergence and be consistent with some recent ESR results~\cite{Khuntia2020}, the gap $\Delta$ is constrained to $g\mu_{\mathrm{B}}B/k_{\mathrm{B}}$ with $g\in[2.14,2.23]$ (except in zero field). We have done a global fit of the $120$ differences allowed by the $16$ datasets shown in Fig.~\ref{fig:main2}, thus employing a total of $17$ global parameters. A few of these fits are shown in Fig.~\ref{fig:main4}(a). The distribution of the gap is only necessary at low and moderate fields, see the inset in Fig.~\ref{fig:main4}(b). We obtain a fraction $f=0.227(1)$ and $\Delta(B=0)=2.06$~K. It is then possible to extract the field-independent contribution $C_{p}^{\star}=C_{p}^{\mathrm{Kago}}+C_{p}^{\mathrm{Phon}}$ which is displayed in Fig.~\ref{fig:main4}(b) for a few fields. The curves are brought to coincide at all fields and on an extended temperature range up to $8$~K showing that the field-dependent character of the total specific heat is well modeled by the Schottky-like contribution only.

\section{\label{sec:anomalies}Field-induced anomalies at very low temperatures}

\begin{figure}\centering
\includegraphics[width=\columnwidth]{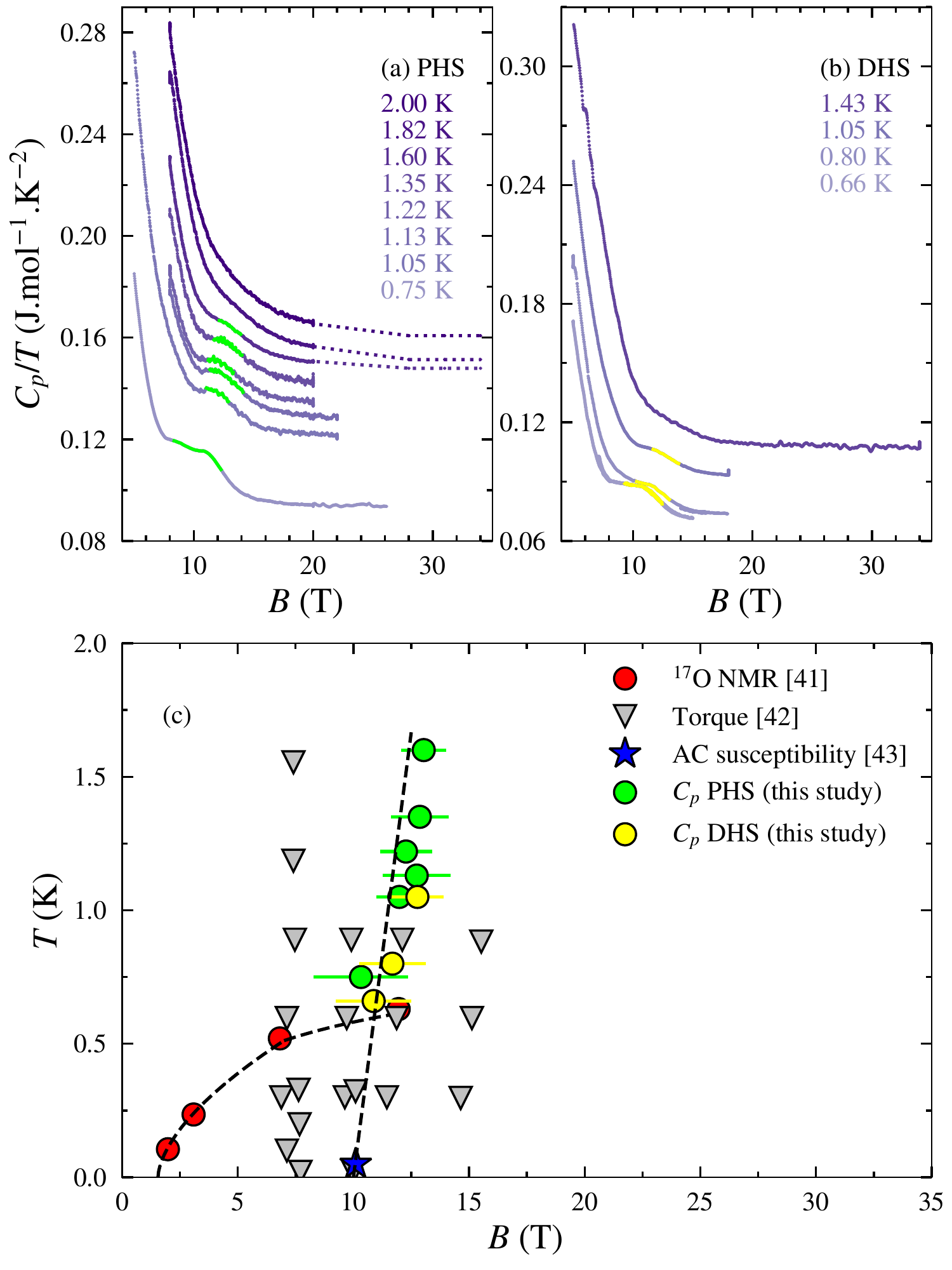}
\caption{\label{fig:main5}(a), resp.~(b) We show field scans obtained at very low temperatures $T\leq2$~K on the PHS single crystal, resp.~the DHS single crystal. Tiny anomalies associated to a minute entropy gain are clearly evidenced at the lowest temperatures, but quickly disappear for temperatures $T\geq1$~K. The parts highlighted in green (PHS), resp.~yellow (DHS), delineate the anomalies (downward concavity). (c) Anomaly positions observed in the latter specific heat measurements are reported in a $B-T$ phase diagram along with anomaly positions detected with other techniques ($^{17}$O NMR spin-lattice relaxation, from Ref.~\cite{Jeong2011}, torque, adapted from Ref.~\cite{Asaba2014} and AC susceptibility, adapted from Ref.~\cite{Helton2009PhD}). The dashed lines are guides for the eyes.}
\end{figure}

We finally focus on the very low-temperature data. In Fig.~\ref{fig:main5}(a) and Fig.~\ref{fig:main5}(b), we show some field scans obtained on the PHS and the DHS single crystals for temperatures $0.7\leq T\leq2$~K. By looking at the lowest temperature data, we clearly distinguish three regimes. At these low temperatures, the low-field specific heat is rapidly decreasing: this is the characteristic Schottky-like contribution. The intermediate-field part displays an anomaly, maybe indicating a field-induced transition. The high-field part is nearly constant, in line with a field-independent kagome contribution $C_{p}^{\mathrm{Kago}}$. As the temperature is increased above $1$~K, we observe a progressive vanishing of the anomaly and see the Schottky-like tail shifting to higher fields. By examining the second derivative of $C_{p}/T$, we are able to delineate the field extent of the field-induced anomaly, which is denoted by a downward concavity~\cite{sup}. The increase in $C_{p}/T$ associated with this anomaly appears to be limited to $25$~mJ.mol$^{-1}$.K$^{-2}$ down to $0.7$~K but we lack the full temperature evolution to determine the precise entropy gain. Such a small field-induced anomaly does not seem compatible with a proper phase transition related to the kagome physics and may reflect a crossover in the defect properties. We draw a tentative $B-T$ phase diagram from the positions extracted and compare our findings with others already reported in the literature, see Fig.~\ref{fig:main5}(c). The reported crossover region has a limited field dependence, similarly to what was observed with torque and ac susceptibility measurements~\cite{Asaba2014,Helton2009PhD}. This crossover is however at variance with the the transition disclosed with $^{17}$O NMR spin-lattice relaxation~\cite{Jeong2011} at lower temperature to a gapped phase. Whatever the origin of these physics, it is restricted to very low temperatures ($T<2$~K) and low to moderate fields ($B<16$~T).

\section{\label{sec:discussion}Discussion}

This discussion section is divided into three parts. In the first part, we compare our findings for the kagome specific heat to existing gapless models. The second part is devoted to possible origins of the Schottky-like contribution. In the third part, we end up examining the $B-T$ phase diagram at very low temperatures.

\subsection{Kagome specific heat}

Our experimental findings bring severe constrains to any model for the ground state and its elementary excitations. Indeed, let us recall that the raw specific heat of herbertsmithite is found to be field-independent for $1\leq T\leq4$~K in fields $28\leq B\leq34$~T ($g\mu_{\mathrm{B}}B\sim J/4)$ and that $C_{p}(T)\propto T^{\alpha}$ with $\alpha\sim1.5$ in the $T\rightarrow 0$ limit (typically for $T\leq1.5$~K). We remind that both the phonon and Schottky-like contributions are negligible for this set of values, then giving access to the kagome specific heat. These experimental facts are  completed by the HTSE$+s(e)$ numerical method to fit the data which also leads to the robust finding of the power law $C_{p}(T)\propto T^{\alpha}$ with $\alpha\sim1.5$ in the $T\rightarrow 0$ limit~\cite{sup}. In addition, this HTSE$+s(e)$ study then yields hardly any variation for the kagome specific heat from $0$ to $34$~T, see Fig.~\ref{fig:main3}. This is consistent with the data after subtraction of a Schottky-like contribution that is dominant in low to moderate fields.

Many of the recent numerical works seem to promote the gapless $U(1)[0,\pi]$ Dirac spin liquid ground state~\cite{Ran2007,Hermele2008,Iqbal2014,Liao2017,Zhu2018,Jiang2019}. For this fermionic spinon mean-field \emph{ansatz}, the low-energy dispersion is linear and isotropic in zero field and pockets form for $B\neq0$ whose radius is proportional to $B$. At the mean-field level, one then expects a change from $C_{p}(T)\propto T^2$ in zero field to $C_{p}(T)\propto BT$ in the $k_{\mathrm{B}}T\ll\mu_{\mathrm{B}}B$ limit~\cite{Ran2007}. Neither the $T$ nor the $B$ dependence are compatible with our experimental results. Moreover, when the spinon Fermi velocity is adapted from the $^{17}$O NMR measurements of the kagome susceptibility in fields $B\leq6.7$~T~\cite{Khuntia2020}, the deduced specific heat magnitude is much smaller than the actual experimental magnitude. The logarithmic correction due to fluctuations of emergent gauge fields derived in Ref.~\cite{Kim1997} for the zero field case results in a shift $\Delta C_{p}(T)\propto T^{2}\ln{(a/T)}$ (where $a\sim J/k_{\mathrm{B}}$). This correction fails to account for the experimental data since it yields $a<1$~K.

In order to better account for the exponent $\alpha\sim1.5$, we tentatively reimagine the fermionic spinon model through the dispersion $e_{\mathbf{k},\pm} \propto |\mathbf{k}|^{2/\alpha}$ featuring subquadratic band touching points ($1<\alpha<2$) rather than Dirac nodes ($\alpha=2$).  This allows to recover $C_{p}(T)\propto T^{\alpha}$ in zero field but also leads to a field dependent linear in $T$ specific heat in large fields, at odds with our findings.

In the vein of a spinon dispersion featuring approximate Dirac cones at low-energy, it has been proposed that the Fermi level does not exactly intersect with the touching points but rather cuts out a small Fermi surface~\cite{Hering2019}. In such a case, one expects a linear specific heat at low-temperature. On the one hand, a fit of our data to such asymptotic behavior yields a small spinon ``Sommerfeld coefficient'': $\gamma\sim50$~mJ.mol$^{-1}$.K$^{-2}$ is an upper bound. On the other hand, although it is not clear how high fields affect such a small density of states, we rely upon our HTSE$+s(e)$ analysis and its extension to the zero-field limit which firmly precludes any linear behavior through the robust prediction of $\alpha\sim1.5$.

A nematic striped spin liquid ground state has been proposed as well, for which the excitation spectrum turns out to be gapless beyond the mean-field description~\cite{Clark2013,Laurita2019,Zhang2020}. Unfortunately, we cannot easily predict what the specific heat would be considering the associated anisotropic spinon transport. This calls for further theoretical developments.

Other scenarios have been suggested but fail to match with the full set of experimental findings. For instance, a $U(1)$ (large) Fermi surface spin liquid ground state~\cite{Ma2008} would provide a simple explanation for the very limited field dependence of the specific heat but is discarded through our small estimate for any spinon ``Sommerfeld coefficient''. Besides, $U(1)$ gauge field fluctuations should result in a correction $\Delta C_{p}(T)\propto T^{2/3}$~\cite{Kim1997}, at odds with our peculiar $\alpha$ value. Moreover, to be stabilized, this state requires a substantial next-nearest neighbor ferromagnetic coupling between kagome spins that is not legitimate for herbertsmithite~\cite{Jeschke2013}. Another example is the proposal of an algebraic vortex liquid~\cite{Ryu2007}, with spinless excitations. Although it is not clear if it would be compatible with our non-trivial power law, it is based on a large easy-plane anisotropy ($XY$ model) that is not in line with the Heisenberg character of herbertsmithite.

To conclude, in the framework of herbertsmithite, on the one hand displaying a kagome-induced spin liquid behavior (down to very low temperatures) but on the other hand not achieving yet the ideal scenario of fully occupied and isolated kagome planes, we are left with no satisfactory physical picture. Let us also mention that, although the series expansion can reproduce our specific heat data, the corresponding computed  susceptibility does not match the small and nearly field-independent susceptibility measured with $^{17}$O NMR for fields $B\leq6.7$~T~\cite{Khuntia2020}. Indeed, the susceptibility at $T=0$ is computed from the field dependence of the ground state energy: $\chi_{0}(B)=-(\partial e_{0}(B)/\partial B)/B$. With the $B^2$ dependence of Eq.~\ref{eq:energy}, this leads to a large field independent $\chi_{0}=3.3\times10^{-3}$~e.m.u.~(mol f.u.)$^{-1}$ more than $10$ times larger than the NMR susceptibility, while for a slightly higher exponent, $B^{\nu}$ with $\nu>2$, compatible with the subquadratic band touching picture ($\nu=\alpha+1$), $\chi_{0}$ vanishes in zero field but increases steeply with the applied field as $\chi_{0}\propto B^{\nu-2}$. However, it may not be suitable to compare directly these two susceptibilities for an inhomogeneous system. Indeed, while the dilution effect is averaged in the computed susceptibility, the NMR results could concentrate on some ``far-from-defect'' copper sites with a susceptibility that is reduced from the uniform susceptibility~\cite{Gregor2008}. Whether this argument is enough to explain such a discrepancy is questionable.

\subsection{Defects in herbertsmithite}

The quantitative analysis of our specific heat data clearly points to a reduced kagome contribution as compared to that expected from HTSE$+s(e)$ results for the pure QKHA model~\cite{Bernu2020}. In our simple approach, this has been addressed through the introduction of $p\sim10$~\% in-plane vacancies in the series computations. This concentration is surprisingly high considering the statement of a full copper occupancy of the kagome lattice that is inferred from cutting edge x-ray scattering and spectroscopy~\cite{Freedman2010,Smaha2020}. In parallel, it has been known for long that herbertsmithite harbors a substantial amount of nearly free magnetic entities through the Curie tail and Schottky-like anomaly observed at low temperatures and low fields. When treated as $S=1/2$ impurities, the Schottky-like anomaly analysis yields a concentration $f\sim23$~\% for the DHS sample. Why the kagome planes are so much impacted calls for an advanced discussion about the various types of defects. We also emphasize that torque, ESR and $^{17}$O NMR uncovered at least two types of defects~\cite{Zorko2017,Khuntia2020}. At low-temperature, while the main part of the NMR spectrum undergoes a large broadening ($\propto1/T$), two additional lines emerge: one for copper sites with a Curie-like susceptibility ($\propto1/T$) and the other for copper sites with an almost null susceptibility.

First sticking to a full copper occupancy in the planes, this leads to a chemical formula Zn$_{1-x}$Cu$_{3+x}$(OH)$_6$Cl$_2$ with $x=f\sim 0.2$. The effective dilution of the kagome planes that we find is only substantiated when some of the kagome sites are somewhat set apart because of the presence of copper ions on the zinc sites. We can no longer simply argue that copper ions on the zinc sites are decoupled and only responsible for the Curie tail and the Schottky-like anomaly at low temperatures and low fields: they could create local perturbations on adjacent triangular plaquettes (in the top and bottom kagome planes), yielding a reduced kagome specific heat. This is in line with our observation that the ratio of $p$ values for the effective dilutions matches the ratio $q$ of Curie constants. Two prospective ideas emerge:
\begin{enumerate}[label=(\roman*)]
\item Copper ions on the zinc sites produce a Jahn-Teller distortion impacting on adjacent triangular plaquettes. In this regard, the local deformations and their influence on the in-plane couplings (preserving the threefold crystal symmetry) have to be quantified. As a note, one does not expect a spin liquid state to be destabilized by some equilateral defect triangles: see the case of the breathing kagome lattice on which the spin liquid state survives for $J_{\bigtriangleup}/J_{\bigtriangledown}$ as low as 0.25~\cite{Orain2017,Repellin2017,Iqbal2018}.
\item A substantial exchange couples in- and out-of-plane copper ions. To our knowledge, this possible interaction has not been estimated to date. It would lead to the formation of spin multimer defects. For instance an effective S=1/2 could arise from the formation of a trimer involving one interlayer Cu and $2$ others from the adjacent kagome planes. How this can be reconciled with an effective decoupled spin $1/2$ picture remains to be investigated.
\end{enumerate}

We can also consider that a minute proportion of zinc ions sits on the kagome sites. This proportion would be bounded by the uncertainty (a few \%) of aforementioned structural studies~\cite{Freedman2010,Smaha2020} and leave room for a still significant amount of copper ions on the zinc sites. In this framework, we can trace back to several theoretical studies of spinless impurities in quantum spin liquids~\cite{Kolezhuk2006,Gregor2008,Rousochatzakis2009,Patil2020}. It has been found that spin textures setting in the vicinity of a spinless impurity, in the form of frozen dimers and orphan spins, are responsible for a uniform Curie-like susceptibility ($\propto1/T$). It would then mix with the susceptibility of the copper ions on the zinc sites.

As observed in most antiferromagnetic correlated systems~\cite{Alloul2009,Wollny2011}, in any of these scenarios, one can expect these defects to be screened by the surrounding spins even leading to a 
Kondo-like effect~\cite{Gomilsek2019} and the perturbation to propagate in the kagome planes. For example, depending on the spin liquid class, it has been shown that spinless impurities can induce a staggered susceptibility~\cite{Kolezhuk2006,Rousochatzakis2009} that would explain the large spectral broadening of the $^{17}$O NMR line observed at low temperatures~\cite{Olariu2008,Khuntia2020}. Alternatively, orphan spins resulting from the effective dilution may form a random singlet state, as set in the different context of a valence bond crystal ground state~\cite{Singh2010}. Yet, the predicted sublinear contribution to the specific heat does not seem compatible with our results.

To make significant progress in the comprehension of defects, it would now be necessary to study a collection of zinc-paratacamite samples among which the zinc to copper ratio is precisely varied at the verge of the ideal ZnCu$_{3}$ stoichiometry. It is also worth comparing with other quantum kagome compounds as in the zinc-barlowite series~\cite{Tustain2020}, which may involve other types of defects.

\subsection{Phases at very low temperatures}

Our field-induced anomalies at very low temperatures add up to other evidences of marginal physics out of a quantum spin liquid regime. These marginal physics show up clearly for temperatures lower than $k_{\mathrm{B}}T\sim J/100$ in fields lower than $B\sim10$~T. Such a field-induced transition from a gapless to a gapped regime was also found in brochantite which arguably is a spin liquid with a spinon Fermi surface~\cite{Gomilsek2017}. Our specific heat data, here, are not complete enough to precisely extract the associated entropy gain but an upper estimate of $17$~mJ.mol$^{-1}$.K$^{-1}$ can be integrated between $600$~mK and $3$~K, which is about $1/350$ of the total entropy for spins $1/2$. As recalled in Sec.~\ref{sec:anomalies}, $^{17}$O NMR~\cite{Jeong2011}, torque~\cite{Asaba2014} and ac susceptibility~\cite{Helton2009PhD} measurements have already evidenced several crossovers at very low temperatures. We can definitely distinguish two crossover categories. So far, the crossover at temperatures $T<1$~K from the dynamical gapless regime to a frozen gapped regime has only been clearly highlighted with NMR relaxation measurements~\cite{Jeong2011}. On top, as pointed out in the present study, a more ``vertical'' crossover region is observed at fields $7<B<16$~T, extending to slightly higher temperatures~\cite{Asaba2014,Helton2009PhD}. While we adopt a simple and conservative method to extract a single ``vertical'' crossover, looking at the concavity change of specific heat curves, torque experiments (which track the anisotropy of the susceptibility) reveal two sets of such crossovers at nearly constant fields by looking at each inflection in the signal derivative~\cite{Asaba2014}.

As discussed in previous works~\cite{Jeong2011,Mendels2016}, low-energy deviations to the pure QKHA model are suspected to be responsible for the crossover from the dynamical gapless regime to the frozen gapped regime. The sizeable Dzyaloshinskii-Moriya (DM) component estimated from ESR measurements~\cite{Zorko2008,ElShawish2010}, $D_{z}\sim0.04-0.08\;J$ and the possible easy-axis exchange anisotropy $\sim 0.1\;J$ deduced from susceptibility measurements~\cite{Han2012} are often invoked. They are the main perturbations, setting in a range $\sim8-19$~K, well above the detected crossovers. In the case of DM interaction, the value for $D_{z}$ is close to but below the predicted quantum critical point in zero field ($D_{z}\sim0.1J$)~\cite{Cepas2008} and the crossover would be explained by a combined effect with the applied field. This reasoning could also be relevant in case of anisotropic symmetric exchange. We also note that a lower limit of a DM instability of 0.012~$J$ has been found recently using a tensor network states approach~\cite{Lee2018}, at $T=0$, and could match better with our findings.

Interestingly, recent second harmonic generation~\cite{Laurita2019}, torque and ESR experiments~\cite{Zorko2017} point to a global symmetry reduction of the kagome lattice at low-temperature. It is attributed to a minute monoclinic distorsion, below the threshold of up-to-date structural refinements. From a combination of infrared spectroscopy measurements and \emph{ab initio} density functional theory calculations, it has been inferred to the strong magnetoelastic coupling in the kagome planes which seems to be enforced upon cooling~\cite{Li2020}. Then, the crossover from the dynamical gapless regime to the frozen gapped regime could be reminiscent of the clear transition observed in monoclinic clinoatacamite~\cite{Zheng2005,Mendels2007}.

An other inspiring channel to interpret all observed crossovers is the building of antiferromagnetic correlations among impurities. This was emphasized by inelastic neutron scattering for energy transfers $\hbar\omega<0.8$~meV ($\sim J/20$)~\cite{Han2016} and may be related to the small Weiss temperature $\theta\sim 1-2$~K deduced from the low-temperature Curie tail in susceptibility~\cite{sup}. Whatever their nature, it is very likely that some nearly free magnetic entities subtly couple to the kagome copper ions, enhancing the dimensionality of the magnetic lattice at the lowest temperatures. This could definitely be a key ingredient in the crossover from the dynamical gapless regime to the frozen gapped regime. As for the more ``vertical'' crossover, it may simply signal the complete polarization of nearly free magnetic entities.

To close this part, we remind that despite their enigmatic origin, all observed crossovers in herbertsmithite occur at very low temperatures, illustrating once again the robustness of the dynamical gapless regime down to temperatures as low as $\sim J/100$ whatever the field intensity.

\section{Conclusions and outlook}

In herbertsmithite, the ground state of the kagome spin system is disordered, dynamical and gapless~\cite{Mendels2007,Han2012-bis,Khuntia2020}. It has essential attributes of a critical spin liquid: in the $T\rightarrow0$ limit, the non-trivial power laws for the spin-lattice relaxation rate~\cite{Khuntia2020}, $1/T_{1}(T)\propto T^{\beta}$ with $\beta\sim1$ measured in a moderate field of about $7$~T, and specific heat, $C_{p}(T)\propto T^{\alpha}$ with $\alpha\sim1.5$, point to algebraic spin-spin correlations. The present study leads to a major improvement in the characterization of elementary excitations. Because of the peculiar exponent and very weak field dependence of the kagome specific heat, we rule out both the $U(1)$ (large) Fermi surface and $U(1)[0,\pi]$ Dirac spin liquid scenarios. Yet, the exact nature of the emergent quasiparticles remains to be elucidated. From the continuum exposed with inelastic neutron scattering~\cite{Han2012-bis}, they are not $\Delta S=1$ objects. Still, the observation of a large (Lorentz force driven) thermal Hall effect would be beneficial to confirm their fractional character~\cite{Katsura2010}. The experimental results presented here pave a delimited way for new theoretical developments.

From the field-induced anomalies reported in the present study and previous work, we have constructed a preliminary $B-T$ phase diagram. It is now to be completed using other techniques in order to secure the two-crossover picture we exhibit, fully characterize the phases stabilized at very low temperatures and understand their origin. For instance, it would make sense to check whether the two crossovers are detected with sound velocity measurements (which are sensitive to minute variations in the magnetoelastic couplings) or thermal transport measurements (which are sensitive to mobile heat carriers only). Special signatures may allow to discriminate between a global spin freezing and the polarization of some impurities.

\emph{Note added.} --- During the reviewing process, we got aware of a new NQR study in herbertsmithite~\cite{Wang2021} which confirms the existence of inequivalent copper sites with long and short relaxation times, observed in previous NMR studies~\cite{Fu2015,Khuntia2020}, and determines their respective dynamics. They hint at the coexistence of gapped and gapless excitations. Altogether with our study, the inhomogeneity observed in the dynamics~\cite{Wang2021} calls for a comprehensive account of defects in herbertsmithite.

\begin{acknowledgments}
The authors acknowledge the support of LNCMI-CNRS, member of the European Magnetic Field Laboratory (EMFL), and French Agence Nationale de la Recherche, under Grant No. ANR-18-CE30-0022 «LINK».
\end{acknowledgments}

\end{document}